\documentclass[iop,apj,numberedappendix]{emulateapj}
\usepackage{amsmath,amssymb,color}
\usepackage{url}
\begin{document}
\title{BLUE SUPERGIANT MODEL FOR ULTRA-LONG GAMMA-RAY BURST WITH SUPERLUMINOUS-SUPERNOVA-LIKE BUMP}

\author{Daisuke Nakauchi$^{1}$, Kazumi Kashiyama$^{2}$, Yudai Suwa$^{3}$, and Takashi Nakamura$^{1}$}

\altaffiltext{1}{Department of Physics, Kyoto University, Oiwake-cho, Kitashirakawa, Sakyo-ku,Kyoto 606-8502, Japan} 
\altaffiltext{2}{Department of Astronomy \& Astrophysics; Department of Physics; Center for Particle \& Gravitational Astrophysics; Pennsylvania State University, University Park, PA 16802}
\altaffiltext{3}{Yukawa Institute for Theoretical Physics, Kyoto University, Oiwake-cho, Kitashirakawa, Sakyo-ku, Kyoto 606-8502, Japan}

\begin{abstract}
Long GRBs (LGRBs) have typical duration of $\sim 30\ {\rm s}$ and some of them are associated with hypernovae, like Type Ic SN 1998bw. Wolf-Rayet stars are the most plausible LGRB progenitors, since the free-fall time of the envelope is consistent with the duration, and the natural outcome of the progenitor is a Type Ic SN.  While a new population of ultra-long GRBs (ULGRBs), GRB 111209A, GRB 101225A, and GRB 121027A, has a duration of $\sim 10^4$ s, two of them are accompanied by superluminous-supernova (SLSN) like bumps, which are $\lesssim 10$ times brighter than typical hypernovae. Wolf-Rayet progenitors cannot explain ULGRBs because of too long duration and too bright SN-like bump.  A blue supergiant (BSG) progenitor model, however, can explain the duration of ULGRBs.  Moreover, SLSN-like bump can be attributed to the so-called cocoon-fireball photospheric emissions (CFPEs). Since a large cocoon is inevitably produced during the relativistic jet piercing though the BSG envelope, this component can be a smoking-gun evidence of BSG model for ULGRBs. In this paper, we examine u, g, r, i, and J-band light curves of three ULGRBs and demonstrate that they can be fitted quite well by our BSG model with the appropriate choices of the jet opening angle and the number density of the ambient gas. In addition, we predict that for 121027A, SLSN-like bump could have been observed for $\sim 20 \mbox{-} 80$ days after the burst.  We also propose that some SLSNe might  be CFPEs of off-axis ULGRBs without visible  prompt emission.
\end{abstract}

 \maketitle

\section{Introduction}
Long gamma-ray bursts (LGRBs), whose durations are longer than 2 s,
are considered to originate from core collapses of massive stars
\citep{1993ApJ...405..273W, 1999ApJ...524..262M}.  In standard
collapsar scenario, a relativistic jet launched from a central engine
(a black-hole (BH) accretion disk or a magnetar) burrows through the
progenitor, and finally punches out.  Then the prompt gamma rays are
produced by some dissipation processes in the jet, although the
mechanism is still uncertain \citep[see][for a recent
  review]{Meszaros_2013}.  So far, some LGRBs were associated with
type Ic supernova (SN)
\citep[e.g.,][]{2011arXiv1104.2274H,2012ApJ...756..190Z,
  2013arXiv1305.6832X}, which might imply that collapsars come from CO
Wolf-Rayet (WR) stars.  They are plausible progenitors from the
theoretical point of view in that they have no hydrogen and helium
envelopes, and thus jets can penetrate them more easily
\citep{2003MNRAS.345..575M}.  Also, the typical observed duration of
LGRBs, $\sim 30 \ \rm s$, is consistent with the accretion timescale
of the WR progenitor envelope onto the central engine.

The above WR model could not be applied to recently discovered
ultra-long GRBs (ULGRBs), GRB 101225A, GRB 111209A, and GRB 121027A
\citep{2013ApJ...766...30G, 2013arXiv1302.2352L}.  They all have
observed durations of $\sim 10^4 \ \rm s$, which are much longer than
the anticipated central engine lifetimes of CO WR collapsars.  There
were a few proposals in the literature for the {\it Christmas burst},
GRB 101225A, such as a tidal disruption of a comet by a galactic
neutron star~\citep{2011Natur.480...69C}, or an extragalactic outburst
triggered by a stellar merger in a neutron-star binary system at $z =
0.33$~\citep{2011Natur.480...72T}.  However, at least the former model
is now unlikely in terms of the energetics after the redshifts of
these bursts are determined; $z = 0.847$ for GRB 101225, $z = 0.677$
for GRB 111209A, and $z = 1.773$ for GRB 121027A
\citep{2013arXiv1302.2352L}.  Other possible non-collapsar-jet
scenarios are SN shock breakouts~\citep[e.g.,][]{2006Natur.442.1008C},
or tidal disruption of a star by the supermassive black hole at a
galactic center \citep{2011Natur.476..421B, 2011Sci...333..203B}.
However, these models may be incompatible with the observed spectra
and light curves (see \S \ref{subsec:other_scenario}).

In the collapsar-jet scenario, a longer duration can be simply
interpreted by fall-back accretion of a progenitor
envelope~\citep{2012MNRAS.419L...1Q, 2013ApJ...767L..36W}, or direct
envelope collapse of a more massive and extended progenitor like blue
supergiant (BSG)~\citep[e.g.,][]{2001ApJ...556L..37M}.  In this paper,
we consider the latter possibility.  At first sight, the extended
envelope seems problematic for the jet penetration, but the accretion
of the massive envelope activates the central engine long enough for
the jet to break out the progenitor.  Indeed, analytical and numerical
calculations showed that relativistic jets from BSG collapsars can
penetrate the progenitor envelopes
\citep[e.g.,][]{2011ApJ...726..107S, 2012ApJ...754...85N} and that the engine activity may
last as long as the observed duration of the ULGRBs
\citep[e.g.,][]{2012ApJ...752...32W}.

In this paper, we focus on the afterglow phase of ULGRBs.  Especially
in GRB 111209A afterglow, a J-band bump with AB mag $\sim 22$ has been
confirmed around $10\mbox{-}50$ days after the burst
\citep{2013arXiv1302.2352L}.  This emission may reflect the
contribution from an underlying SN and is an unique signature of
ULGRBs.  This underlying SN component is up to $\lesssim 10$ times
brighter than hypernovae associated with LGRBs and is comparable to
the so-called superluminous supernovae (SLSNe).  We investigate the
possibility that such a SLSN-like component is ascribed to a {\it
  cocoon-fireball photospheric emission}
(CFPE)~\citep{Kashiyama_et_al_2013}.  In our previous work, we already
proposed the idea that such a CFPE may be a necessary counterpart of
BSG GRBs and estimated its basic properties using a simple analytical
model.  In this paper, we refine our model of the CFPE and explicitly
show that the underlying SN component of GRB 111209A afterglow is quite well
explained by the CFPE.
We also apply our model to the other two bursts, GRB 101225A and GRB 121027A.
Then we suggest that some fraction of the
observed SLSNe might be originated from the CFPE of off-axis ULGRBs.
CFPEs may become a smoking gun of the BSG model for ULGRBs.

This paper is organized as follows. In \S \ref{sec:ULGRB}, we review
the observational results of ULGRBs. In \S \ref{sec:BSG}, we propose
the BSG model for those ULGRBs. In \S \ref{sec:cocoon}, we explicitly
show that the SLSN-like bump of GRB 111209A afterglow can be fitted
quite well by the CFPE model.  We also apply our model to the other
two bursts, GRB 101225A and GRB 121027A.  \S\ref{sec:diss} and
\S\ref{sec:sum} are devoted to the discussions and summary.  In
Appendix \ref{app}, we describe our model and the calculation method
in detail.

\section{Ultra-Long Gamma-Ray Bursts}\label{sec:ULGRB}
In this section, we briefly summarize the observed features of
recently detected ULGRBs; GRB 101225A, GRB 111209A, and GRB 121027A.
We show some representative features of them in Table
\ref{tab:obs}~\citep[for details, see][]{2013ApJ...766...30G,
  2013arXiv1302.2352L}.  They all have ultra-long durations of $\sim
10^{4} \ \rm s$ and isotropic energies of $E_{\gamma, \rm iso} \sim
10^{52\mbox{-}53} \ \rm erg$ in prompt
phase.\footnote{\cite{2013ApJ...766...30G} showed that the estimated
  spectrum peak energy $E_{\rm p}$ and $E_{\gamma, \rm iso}$ of GRB
  111209A agree with the $E_{\rm p}-E_{\gamma, \rm iso}$ correlation
  \citep{2002AA...390...81A} within $2\ \sigma$ level. While both
  prompt emission and X-ray flare of GRB 121027A satisfy the $E_{\rm
    p}-L_{\rm p}$ correlation \citep{2004ApJ...609..935Y}, the $E_{\rm
    p}-E_{\gamma, \rm iso}$ correlation holds only for the prompt
  emission \citep{2013arXiv1302.4876P}. Physical values of GRB 101225A
  were not determined well and the relationship with these empirical
  correlations are uncertain.}
  
X-ray afterglows were observed by {\it Swift} X-Ray Telescope (XRT).
For GRB 111209A and GRB 121027A, the light curve shapes are similar to
those of conventional LGRBs, i.e., consisting of a steep decay, a
shallow decay and a normal decay phase~\citep{2006ApJ...642..354Z}.
The jet breaks have not been confirmed until $\sim 2 \times 10^6\ {\rm
  s}$, which gives constraints on the jet opening angles; $\theta_{\rm
  j} > 12^{\circ}$ (GRB 111209A) and $\theta_{\rm j} > 10^{\circ}$
(GRB 121027A) \citep{2013arXiv1302.2352L}.  For GRB 101225A, X-ray
afterglow is not detected after the end of the steep decay phase $\sim
10^5 \ \rm s$ and no constraint is given to the jet opening angle
\citep{2013arXiv1302.2352L}.

A remarkable feature was found in the UV/optical/IR afterglow of GRB
111209A~\citep{2013arXiv1302.2352L}, which we mainly discuss in this
paper.\footnote{Even before $t_{\rm obs} \lesssim 1\ {\rm day}$, some
  complex features (rebrightening and decay) are seen in the
  UV/optical/IR light curves. \cite{2013arXiv1306.1699S} performed
  detailed temporal and spectral analysis of these features. On the
  contrary, we focus on the afterglow light curves after $t_{\rm obs}
  \gtrsim 1\ {\rm day}$.}  For $t_{\rm obs} \gtrsim 1$ day, the u-band
light curve exhibits a single power law decay with $t^{-1.38}$, which
is similar to X-ray band, while the J-band light curve decays more
slowly as $t^{-0.5}$.  The difference between u-band and J-band
temporal indices ($\sim 0.9$) contradicts that predicted from the
standard external shock model ($0.25$) which can be seen from
Eq. \eqref{eq:sp_light_curve}.  \cite{2013arXiv1302.2352L} argued that
these emissions may be supernova bump, which are clearly different
from other GRB-associated SN, and are as bright as the so called
superluminous supernovae (SLSNe), which are typically brighter than the normal SN by a factor of $\sim$ 100 \citep[e.g.,][]{2012Sci...337..927G}.
GRB 101225A also showed very
shallow temporal decay ($\propto t^{-0.59}$ in r-band and $\propto
t^{-0.34}$ in i-band) for late-time GRB afterglow.  These are also
ascribed to an underlying supernova component whose peak luminosity is
comparable to the GRB-associated hypernova, SN 1998bw
\citep{2013arXiv1302.2352L}.  Although such a SN-like bump has not
been reported yet for GRB 121027A, it may be an unique feature of
ULGRBs.

\begin{table}[!htb]
\caption{Observed Characteristics of ULGRBs. } 
\begin{center}
{\begin{tabular}{lcccc}
\hline
& 111209A  & 101225A  & 121027A \\ \hline
$E_{\rm iso}^{\rm obs}$ $(10^{53}\ {\rm erg})$  & $5.8$$^{\dagger}$ & $\gtrsim 0.12$$^{\ddagger}$ & $2.0^{\diamondsuit}$ \\
$\delta t_\gamma^{\rm obs}$$({\rm s})$  & {15000}$^{\dagger}$ & $\gtrsim 2000^{\dagger}$ &   $10000^{\diamondsuit}$ \\ 
$\theta_{\rm j}$  & $\gtrsim 12^{\circ}$$^{\ddagger}$ &  $-$   &  $\gtrsim 10^{\circ}$$^{\ddagger}$ \\
$z$  \ & $0.677^{\dagger}$  & $0.847^{\ddagger}$  & $1.773^{\ddagger}$ \\
\hline
\end{tabular}}
\end{center}
{\bf Notes.} For GRB 101225A, the prompt emission was already active
when {\it Swift} BAT first slewed to the burst location
\citep{2013arXiv1302.2352L}, which gives lower limits.  GRB 121027A
can be divided into the prompt emission with the duration of $\sim
200\ {\rm s}$, and the giant X-ray flare lasting for $\sim 10^4\ {\rm
  s}$~\citep{2013arXiv1302.4876P}.  Here, we include the giant X-ray
flare to the prompt phase, given that both of them can be originated
from the central engine.

{\bf Reference.} 

$^{\dagger}$\cite{2013ApJ...766...30G}, $^{\ddagger}$\cite{2013arXiv1302.2352L}, $^{\diamondsuit}$\cite{2013arXiv1302.4876P}
\label{tab:obs}
\end{table}

\section{Blue Supergiant Model}\label{sec:BSG}

\begin{figure*}[!htb]
\begin{center}
\includegraphics[scale=0.25]{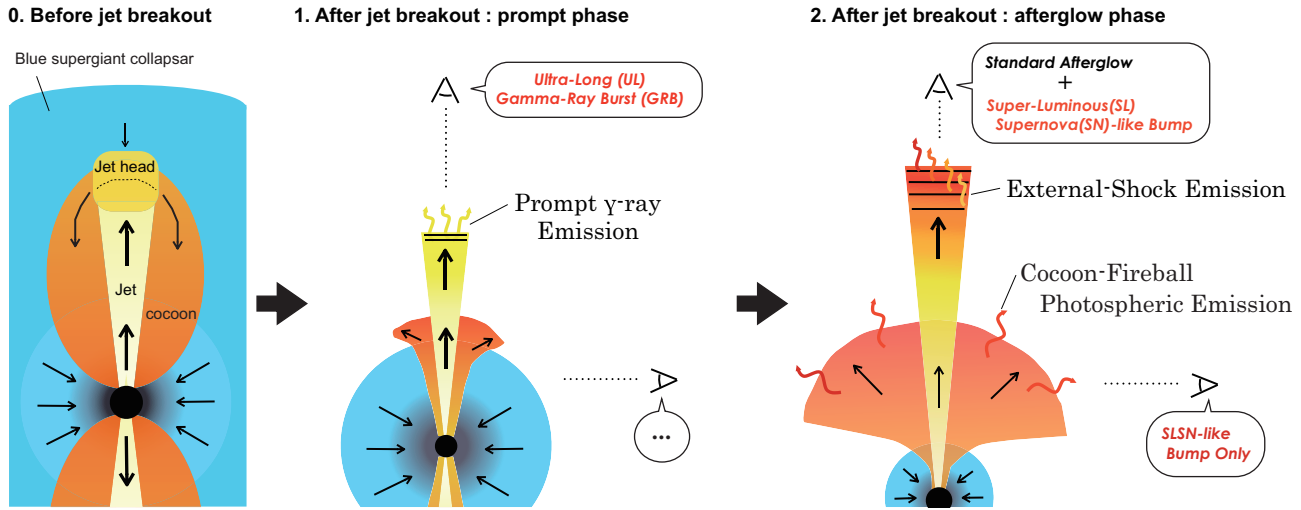}
\caption{The schematic picture of our BSG~(blue supergiant) model for ULGRBs~(ultra-long gamma-ray burst).
0) Before jet breakout.
A jet from the central engine is piercing through a BSG star. 
The jet head moves with non-relativistic speed forming a hot cocoon.
The activity of the central engine continues as far as the envelope matter falls onto the central black hole.
1) After jet break out: prompt phase. If we observe the event along the jet axis, it is a ULGRB with the duration of $\sim 10^4$ s. From the off-axis direction, we see nothing or dim orphan emissions.
2) After jet break out: afterglow phase.
Along the jet axis we see the standard afterglow emissions and the CFPE~(cocoon-fireball photospheric emission) as a SLSN~(superluminous supernova)-like bump.
From the off-axis direction, we see only a SLSN-like bump.}
\label{fig:BSGmodel}
\end{center}
\end{figure*}

In this section, we review the BSG model for ULGRBs. It should be
emphasized that this model naturally exhibits the ultra-long duration
of these events. In addition, it is predicted that this model
accompanies a SLSN-like bright bump in the afterglow phase, which
might be actually observed in the optical/infrared band.  
In this paper, we refer the pre-supernova models to \cite{2002RvMP...74.1015W}\footnote{In our previous study~\citep{2012ApJ...759..128N}, we defined the point where the density becomes $10^{-7}\ {\rm g}\ {\rm cm}^{-3}$ as the radius for jet breakout, for stellar models with metallicity of $10^{-4}$ Z$_{\odot}$.
Since each stellar model in \cite{2002RvMP...74.1015W} has an outermost density which is different from each other, we adjusted them to the same value.
In this paper, following them, we adopted the point of $10^{-7}\ {\rm g}\ {\rm cm}^{-3}$ as the effective stellar surface in calculating the jet propagation.}, whose data are available from the webpage of Alex Heger\footnote{\url{http://2sn.org/stellarevolution/data.shtml}}.
The schematic picture of our model is shown in Fig. \ref{fig:BSGmodel}.  We describe
 details of  our model below.

Our calculations are based on the collapsar scenario of LGRBs
\citep{1993ApJ...405..273W, 1999ApJ...524..262M}.  In this scenario,
ultra-relativistic jets are launched  from the BH accretion disk system
which is formed following massive progenitor collapses so that the  GRB
prompt emission is  raised after jets break out the
progenitor envelope.  The central engine is essentially kept active
while the progenitor envelope can be accreted onto it
\citep{2008MNRAS.388.1729K} with a typical time scale of
\begin{equation}
\delta t_\gamma \sim t_{\rm ff}(R_{\ast}) (1+z)
 \sim 77
\left(\frac{R_{\ast}}{R_{\rm WR}}\right)^{3/2}
\left(\frac{M_{\ast}}{M_{\rm WR}}\right)^{-1/2} (1+z)
\ {\rm s},
\label{eq:t_acc}
\end{equation}
where $t_{\rm ff}(r) = \sqrt{r^3/GM_r}$ is the free-fall timescale of
a mass shell at mass coordinate $M_r$ and radius $r$.  For a WR
progenitor with radius $R_{\rm WR} \sim 2 \times 10^{10}\ {\rm cm}$
and mass $M_{\rm WR} \sim 10\ M_{\odot}$~\citep{2002RvMP...74.1015W},
typical durations of observed LGRBs are reproduced from Eq.(1).  On the other
hand, in order to explain the durations of ULGRBs, one needs to invoke
an envelope accretion of a more massive star with  an extended progenitor
like a blue supergiant~(BSG).  For example, given a BSG with $R_{\ast}
\sim 10^{12}\ {\rm cm}$ and $M_{\ast} \sim 50\ M_{\odot}$, we can
estimate the  accretion time of the envelope as
\begin{equation}
\delta t_\gamma \sim 1.2\times 10^4 
\left(\frac{R_{\ast}}{10^{12}\ {\rm cm}}\right)^{3/2}
\left(\frac{M_{\ast}}{50\ M_{\odot}}\right)^{-1/2} (1+z)
\ {\rm s}.
\label{eq:t_accBSG}
\end{equation}
Based on this estimate, BSGs were proposed as progenitors of
ULGRBs~\citep{2013ApJ...766...30G, Kashiyama_et_al_2013}.  According
to the stellar evolution theory, metal poor stars end as BSGs with
massive hydrogen envelopes and typical radii of $10^{12}-10^{13}\ {\rm
  cm}$.  This is because the low opacity envelope suppresses the
line-driven mass loss from the stellar
surface~\citep{2002RvMP...74.1015W}.  Metal poor
stars with mass of $40-140\ M_{\odot}$ are considered to form black
holes with little mass ejection \citep{2003ApJ...591..288H}.  Thus, by
considering a metal-poor BSG like collapsar, we can expect that the very long duration of
ULGRBs are natural outcomes \citep[e.g.,][]{2012ApJ...752...32W}.

Some LGRB afterglow light curves show bumps in UV/optical/IR band,
which imply the presence of underlying SNe
\citep{2006ARA&A..44..507W}.  So far, nearly a dozen of LGRBs are
confirmed to accompany type Ic supernovae, which implies WR stars as
progenitors. Thus UV/optical/IR afterglow provides us a key to
determine the progenitor of GRBs.  As we saw in the previous section,
GRB 111209A afterglow shows a J-band bump and the accompanying
SLSN-like component is suggested.  This association, however, seems
problematic for the BSG model, because SN explosions may fail for
massive BSGs. If BSGs succeeded in SN explosions in some way, this time,
 a large fraction of the progenitor envelope would be ejected away to suppress the long
lasting accretion onto the central engine. This dilemma forced us to consider
another mechanism of SLSN-like bump, which is CFPE~(cocoon-fireball photospheric emission).

In this paper, by considering jet-cocoon structure in the collapsar
envelope, we interpret the SLSN-like bump as the CFPE, which 
 strengthens our previous proposal  that the progenitor of a ULGRB is a BSG-like star \citep{Kashiyama_et_al_2013}.
While a collapsar jet is piercing
through the progenitor envelope, jet energy is dissipated at the jet
head and is stored in a cocoon consisting of hot
plasma~\citep{2003MNRAS.345..575M}.  Along with the jet breakout,
cocoon also breaks out the progenitor and its dynamics can be regarded
as a non-relativistic fireball.  \cite{Kashiyama_et_al_2013} modeled
the photospheric emission from an expanding cocoon fireball in a rather
simple manner. We here refine our model so as to explain the observed
features. To make the paper easier to read for non-experts, we show  the details of numerical methods  in the Appendix.

\section{Cocoon-fireball Photospheric Emission as a Superluminous-Supernova-like Bump of Ultra-long Gamma-ray Burst}\label{sec:cocoon}
In this section, we first focus on the SLSN-like bump of GRB 111209A.
Based on the BSG model, we show that such a bright UV/optical/IR
emission can be interpreted as the cocoon-fireball photospheric
emission (CFPE).  Then, we apply our model to other two events, and
discuss the possible constraints.

The calculation methods we use are shown in Appendix in detail.
The schematic picture of our model is shown in Fig. \ref{fig:BSGmodel}.
Below, we overview our methods;

\begin{itemize}
\item First, we give the progenitor model, and the jet opening angle
  $\theta_{\rm j}$.  Then, for a fixed jet efficiency $\eta_{\rm j}$,
  we calculate the jet propagation within the progenitor envelope, and
  determine the jet breakout time $t_{\rm bo}$ based on our previous
  works \citep{2011ApJ...726..107S,2012ApJ...759..128N,Kashiyama_et_al_2013}
  .  By calculating the mass accretion rate after breakout, we determine the duration of the
  prompt emission $\delta t_{\gamma}$. $\eta_{\rm j}$ is adjusted so as to
  reproduce the observed duration $\delta t_{\gamma}^{\rm obs}$. Cocoon parameters at breakout, $E_{\rm c}(t_{\rm bo})$ and $M_{\rm
    c}(t_{\rm bo})$ are obtained at this stage (see \S
  \ref{subsec:jet} and \S \ref{subsec:prompt}).

\item Second, by considering the jet luminosity after breakout, we
  determine the isotropic energy $E_{\rm iso}$ in prompt phase.  A
  constant radiation efficiency $\epsilon_\gamma$ is introduced and is
  adjusted so as to reproduce the observed $E_{\rm iso}^{\rm obs}$.
  Thus, $\eta_{\rm j}$ and $\epsilon_\gamma$ are chosen appropriately
  for a given $\theta_{\rm j}$ and progenitor model.  The kinetic
  energy of the relativistic ejecta $E_{\rm kin}$ is obtained
  at this stage (see \S \ref{subsec:prompt}).

\item Third, we calculate the afterglow emissions following the
  standard external shock model in \cite{1998ApJ...497L..17S}, and
  compare our results with the observed X-ray light curves (LCs).  To
  begin with, from the X-ray LC slope, the power law index of the
  accelerated electron's energy spectrum $p$ is determined.  Then, the
  electron acceleration efficiency $\epsilon_e$ is strongly
  constrained from the observed X-ray flux, due to the large
  dependence of the theoretical flux on $\epsilon_e$, i.e.,
  $F_{(0.3\mbox{-}10\ {\rm keV})} \propto \epsilon_{e}^{3/2}
  \epsilon_{B}^{1/8} n^{0}$ (see \S \ref{subsec:afg} and
  Eq. \ref{eq:sp_light_curve} with $p=2.5$).

\item Fourth, we calculate the CFPEs, using $E_{\rm c}(t_{\rm bo})$ and
  $M_{\rm c}(t_{\rm bo})$ as the initial parameters of the cocoon
  fireball.  We suppose that the CFPE contributes dominantly to the
  optical/IR bump.  CFPE is attenuated by the host galaxy, and the
  V-band extinction in host galaxy, $A_{\rm V}^{\rm host}$, is
  adjusted so as to reproduce the observed SN-like bump in
  optical/IR band (see \S \ref{subsec:cocoon}).

\item Finally, we suppose that the UV flux is dominantly contributed
  from the external shock emission.  For typical parameters, UV flux
  can be calculated from $F_{\rm UV} \propto \epsilon_{e}^{3/2}
  \epsilon_{B}^{7/8} n^{1/2}$ (see Eq. \ref{eq:sp_light_curve} with
  $p=2.5$).  From the observed UV flux, the appropriate value of the
  magnetic field amplification efficiency $\epsilon_{B}$ is
  obtained for a given ambient gas density $n$.  Thus, if we give the
  ambient gas density $n$, afterglow parameters, $\epsilon_e$,
  $\epsilon_{B}$, and $p$ are set appropriately from the X-ray and UV
  observations (see \S \ref{subsec:afg}).

\end{itemize}

In summary, we have 6 constraints from the observations against 8
unknown parameters.  As the two free parameters, we choose $\theta_{\rm j}$
 and the ambient gas density $n$. In this study, as for the progenitor, we adopt the
BSG model with zero age main sequence (ZAMS) mass of $75\ M_{\odot}$ and
metallicity of $10^{-4}\ Z_{\odot}$ calculated by
\cite{2002RvMP...74.1015W}.\footnote{The observations suggested that
  the ULGRB host galaxies have sub-solar metallicities, and that as
  long as we consider BSG progenitors, they might be originated from
  massive star binary systems rather than single low metal massive
  stars \citep{2013arXiv1306.1699S}.  However the stellar structure is
  not so different in BSGs, so that our progenitor can be regarded as
  a representative model applicable for any scenario, i.e., either the
  consequence of single star evolution or binary evolution.}  In the
pre-collapse phase, this BSG has mass $M_{\ast} \sim 75\ M_{\odot}$
and radius $R_{\ast} \sim 8.6 \times 10^{12}\ {\rm cm}$.  If we give
the jet opening angle $\theta_{\rm j}$ and the ambient gas density
$n$, we can calculate all the features of ULGRBs both in prompt and
afterglow phases.  The model parameters, which fit the observational
data, are summarized in Table \ref{tab:parameter}.

\begin{table}[!htb]
\caption{Model parameters and calculated quantities with zero age main sequence (ZAMS) mass of $75\ M_{\odot}$ and
metallicity of $10^{-4}\ Z_{\odot}$ calculated by
\cite{2002RvMP...74.1015W}  }
\begin{center}
{\begin{tabular}{lcccc}
\hline
                                                         &   111209A                      & 101225A                     & 121027A  \\ \hline
$\theta_{\rm j}$                                &  $12^{\circ}$                   & $12^{\circ}$                &$12^{\circ}$                     \\ 
$\eta_{\rm j}$                                   &  $1.24 \times 10^{-3}$    & $6.2 \times 10^{-4}$   &$1.24 \times 10^{-3}$  \\ 
$\epsilon_{\gamma}$                                           &  $0.38$       & $0.7$                         &$0.1$    \\ 
$p$                                                                       &  $2.5$         & $2.1$                         &$2.6$    \\  
$\epsilon_{e}$                                  &  $0.01$                           & $5 \times 10^{-4}$     &$0.05$    \\ 
$\epsilon_{B}$                                 &  $1 \times 10^{-3}$         & $0.05$                        &$8 \times 10^{-4}$    \\ 
$n$ (${\rm cm}^{-3}$)                                          &  $0.04$       & $0.1$                          &$0.01$    \\ 
$A_{\rm V}^{\rm host}$ (mag)                             &  $0.26$         & $0.58$                          &$1.1$    \\\hline                     
$E_{\rm iso}$ $(10^{53}\ {\rm erg})$    &  $5.9$         & $2.4$                        &$1.5$    \\ 
$\delta t_\gamma$ $({\rm s})$                         &  $9000$          & $5100$                       &$15000$    \\
$E_{\rm kin}$ $(10^{53}\ {\rm erg})$                    &  $9.6$         & $1.0$                       &$14$    \\ 
$E_{\rm c}(t_{\rm bo})$ $(10^{53}\ {\rm erg})$     &  $1.0$         & $0.56$                        &$1.0$    \\ 
$M_{\rm c}(t_{\rm bo})$ $(M_{\odot})$                 &  $5.8$         & $7.0$                          &$5.8$    \\  
 \hline
\end{tabular}}
\end{center}
\label{tab:parameter}
\end{table}

\subsection{GRB 111209A}\label{subsec:111209A}

\begin{figure*}[!tb]
\begin{center}
\begin{tabular}{ll}
\resizebox{80mm}{!}{\includegraphics[]{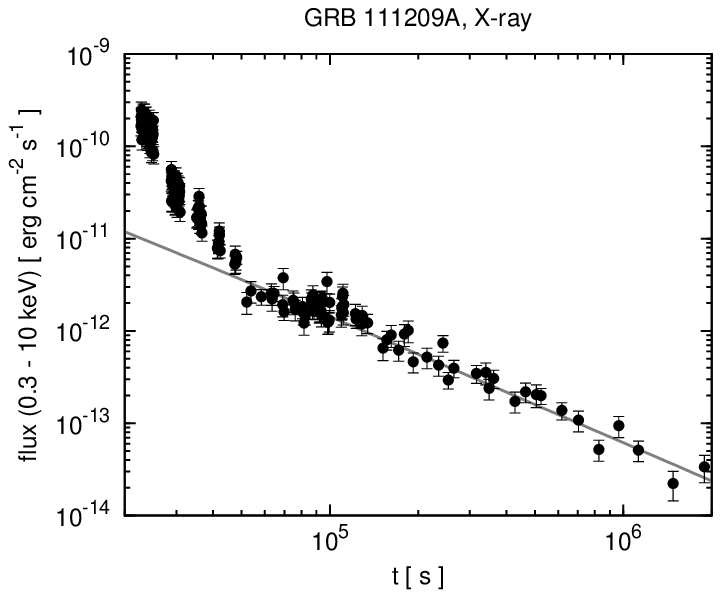}} &
\resizebox{80mm}{!}{\includegraphics[]{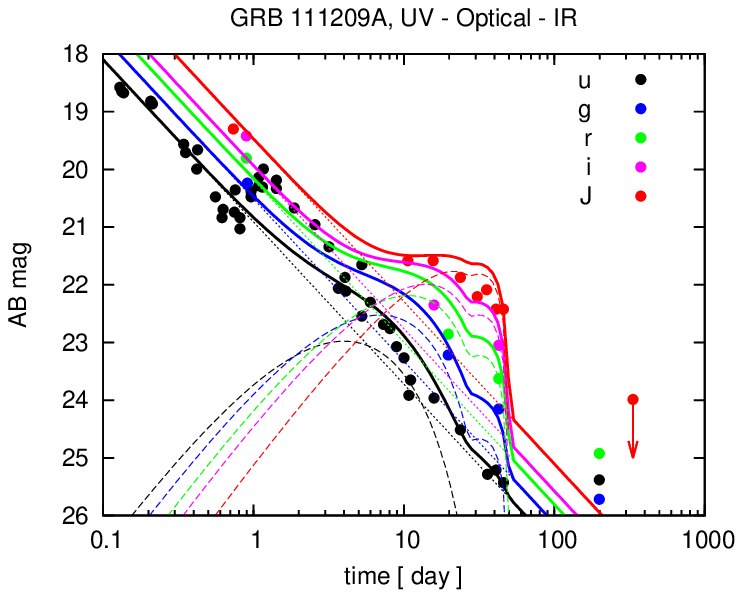}} \\
\end{tabular}
\caption{Theoretical fitting of GRB 111209A afterglow light curves
  (LCs).  The left panel shows the LC in XRT range and the right one
  is in UV/optical/IR range, where the points represent the observed
  data, and the solid lines correspond to the theoretical model in X
  (grey), u (black), g (blue), r (green), i (magenta) and J (red)
  bands, respectively.  While the observations are shown as 
  filled circles with the same colors, respectively. 
  We use a BSG progenitor with zero age main sequence (ZAMS) mass of $75\ M_{\odot}$
  and $10^{-4}\ Z_{\odot}$.  We find that if we give $\theta_{\rm j} =
  12^{\circ}$ and $n = 0.04\ {\rm cm}^{-3}$, other parameters are
  determined from the observations as $\eta_{\rm j} = 1.24 \times
  10^{-3}$, $\epsilon_{\gamma} = 0.38$, $p = 2.5$, $\epsilon_e =
  0.01$, $\epsilon_B = 1 \times 10^{-3}$, and $A_{\rm V}^{\rm host} =
  0.26$ (see Table \ref{tab:parameter}).  In the right panel, the
  thin-dotted lines correspond to the external shock components.  The
  SLSN-like bump, which dominates at later phase, is well reproduced
  by the CFPE (thin-dashed lines) with $E_{\rm c}(t_{\rm bo}) = 1.0
  \times 10^{53}\ {\rm erg}$ and $M_{\rm c}(t_{\rm bo}) =
  5.8\ M_{\odot}$.  We see that the theoretical curves reproduce the observations quite well.
  In the right panel, the error bars are smaller than the data points in $J$, $i$, $r$, and $g$-bands.
But in $u$-band, some error bars are larger than the point size, and are $\pm 0.5$ mag at most.
  The data points at $\sim 200$ day may reflect the
  emissions from the host galaxy \citep[see][for details]{2013arXiv1302.2352L}. }
\label{fig:111209A}
\end{center}
\end{figure*}

For GRB 111209A, we set the jet opening angle as $\theta_{\rm j}
=12^{\circ}$, which is the lower limit given by the non-detection of
the jet break in the X-ray afterglow~\citep{2013arXiv1302.2352L}.  In this case, the observed
isotropic energy $E_{\rm iso}^{\rm obs} \sim 5.8 \times 10^{53}\ {\rm
  erg}$ and the duration $\delta
t_\gamma^{\rm obs} \sim 15000\ \rm s$~\citep{2013ApJ...766...30G} are substantially reproduced by setting the
jet efficiency and the radiation efficiency as $\eta_{\rm j} = 1.24
\times 10^{-3}$ and $\epsilon_\gamma = 0.38$, respectively.  The above
set of parameters $(\theta_{\rm j}, \eta_{\rm j}, \epsilon_\gamma)$
determines the kinetic energy of the relativistic ejecta as $E_{\rm
  kin} = 9.6 \times 10^{53}\ {\rm erg}$, and the internal energy and
the baryon mass of the cocoon as $E_{\rm c}(t_{\rm bo}) = 1.0 \times
10^{53}\ {\rm erg}$ and $M_{\rm c}(t_{\rm bo}) = 5.8\ M_{\odot}$,
respectively (see Table \ref{tab:parameter}).

The left panel of Fig. \ref{fig:111209A} shows the XRT afterglow LC (black dots with error bars)
and the theoretical fitting (solid gray line). The right panel of this figure
shows the UV/optical/IR afterglow LCs and the
theoretical fittings for u (black), g (blue), r (green), i (magenta),
and J (red) bands.  The data points at $\sim 200\ {\rm day}$ may
reflect the host galaxy contribution \citep{2013arXiv1302.2352L}.  We
focus on the normal decay phase starting at $t_{\rm obs} \sim
10^5\ {\rm s}$.  The observed X-ray flux decays as $t^{-1.36}$ for $t
\gtrsim 10^5\ {\rm s}$ \citep{2013arXiv1302.2352L}.  This gives the
power law index of non-thermal electrons as $p = 2.5$.
The X-ray and u-band fluxes are reproduced by setting the ambient gas
density, the electron acceleration efficiency, and the magnetic field
amplification efficiency as $n = 0.04\ {\rm cm}^{-3}$, $\epsilon_e =
0.01$ and $\epsilon_B = 1 \times10^{-3}$, respectively.  One can see
that the standard-afterglow components (thin-dotted lines) roughly
illustrate the observed data for $1\ \rm day \lesssim t_{\rm obs}
\lesssim 5 \ \rm day$,\footnote{One can see that there is a
  re-brightening in u-band at $\sim 1\ {\rm day}$, which also cannot
  be explained by the standard external shock model.  Our target here
  is, however, the SLSN-like component emerged after $\sim 10\ {\rm
    day}$.  So, for simplicity, we treat the power law component of
  the afterglow within the standard model.  Detailed theoretical
  interpretations of this re-brightening are discussed in
  \cite{2013arXiv1306.1699S}.} and the SLSN-like bump dominates in
optical/IR bands for $t_{\rm obs} \gtrsim 5 \ \rm day$.  We find that
by setting $A_{\rm V}^{\rm host} = 0.26$ mag, the CFPEs (thin-dashed
lines) explain the SLSN-like bump quite well.

The model parameters for the above fittings have reasonable values
(see Table \ref{tab:parameter}).  Thus, we can conclude that ULGRB
111209A and the accompanying SLSN-like bump are well reproduced by the
BSG collapsar model.  Note that since the CFPEs are calculated based
on a TypeIIP SN model, the observed bump may be able to be explained by a SN
ejecta, not by a cocoon.  However, a significantly large explosion energy of $\sim
10^{53} \ \rm erg$ ($\sim$ a third of the binding energy of the 
neutron star) is still necessary, which would be very difficult as far
as we consider standard spherical explosions.
 
\subsection{GRB 101225A}

\begin{figure*}[!tb]
\begin{center}
\begin{tabular}{ll}
\resizebox{80mm}{!}{\includegraphics[]{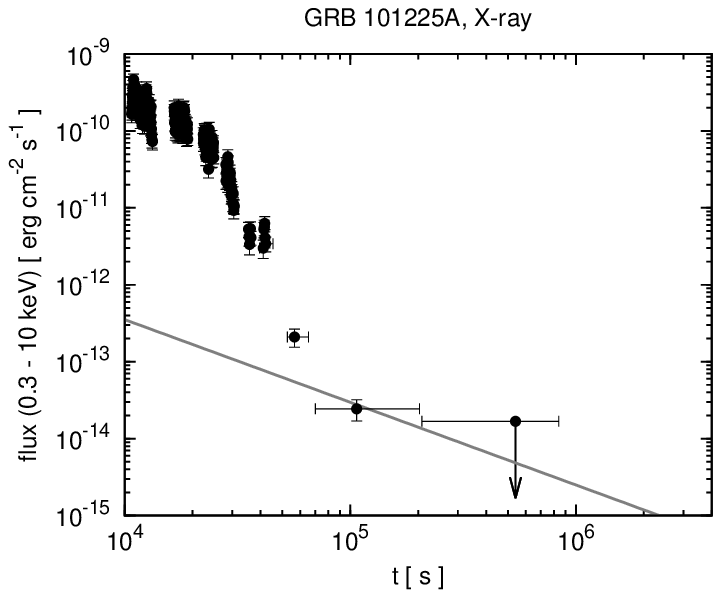}} &
\resizebox{80mm}{!}{\includegraphics[]{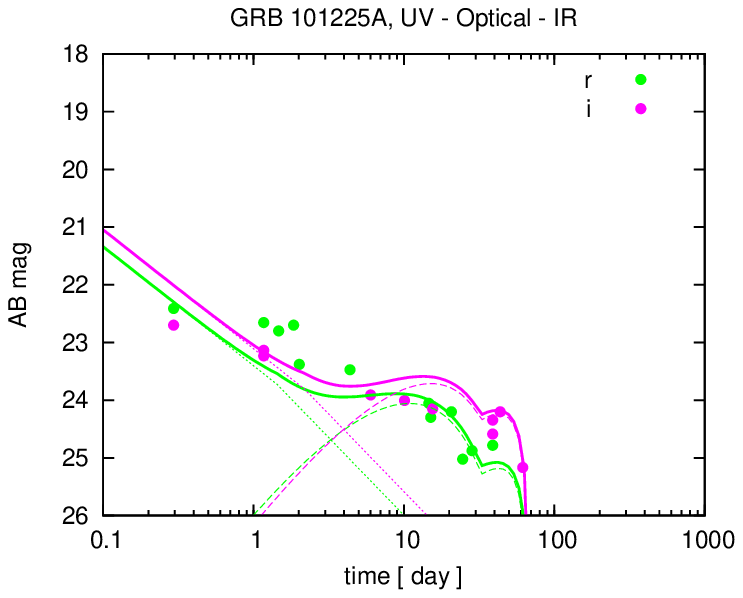}} \\
\end{tabular}
\caption{Same as Fig. \ref{fig:111209A}, but for GRB 101225A.  Despite
  large uncertainties, the observed features are consistent with the
  BSG model, where the external shock components dominate at earlier
  phase (thin-dotted lines) and the CFPE dominates at later phase
  (thin-dashed lines) with $E_{\rm c}(t_{\rm bo}) = 5.6 \times
  10^{52}\ {\rm erg}$ and $M_{\rm c}(t_{\rm bo}) =
  7.0\ M_{\odot}$. Here, the progenitor model is the same as
  Fig. \ref{fig:111209A}.  We have only two observational constraints in r(green) and i(magenta) bands
  . Nevertheless  for a given set of parameters such as $\theta_{\rm j} = 12^{\circ}$,
  $\eta_{\rm j} = 6.2 \times 10^{-4}$, $\epsilon_{\gamma} = 0.7$, $p =
  2.1$, $\epsilon_e = 5 \times 10^{-4}$, and $n = 0.1\ {\rm cm}^{-3}$,
  the remaining parameters are determined from the observations as
  $\epsilon_B = 0.05$ and $A_{\rm V}^{\rm host} = 0.58$ (see Table
  \ref{tab:parameter}). The theoretical curves reproduce the observations rather well.
  In the right panel, the error bars are less than $\pm 0.25$ mag in each band~\citep[see][for details]{2013arXiv1302.2352L}. }
\label{fig:101225A}
\end{center}
\end{figure*}

For GRB 101225A, it is relatively hard to constrain our model
parameters, since we only have a lower limit to the duration and the
isotropic energy of the prompt emission, and no constraint is given to
the opening angle.  Here, we assume the same opening angle
$\theta_{\rm j} = 12^{\circ}$ as GRB 111209A, and take fiducial values
for the jet efficiency, $\eta_{\rm j} = 6.2 \times 10^{-4}$ and the
radiation efficiency, $\epsilon_{\gamma} = 0.7$.  The parameter set
gives $\delta t_\gamma \sim 5100 \ \rm s$ and $E_{\rm iso} = 2.4
\times 10^{53} \ \rm erg$, which exceed the observed lower limits, and
$E_{\rm kin} = 1.0 \times 10^{53} \ \rm erg$ is also obtained.  Cocoon parameters are
also calculated as $E_{\rm c}(t_{\rm bo}) = 5.6 \times 10^{52}\ {\rm
  erg}$ and $M_{\rm c}(t_{\rm bo}) = 7.0\ M_{\odot}$, respectively.

The left panel of Fig. \ref{fig:101225A} shows the afterglow LC in XRT
band.  For $t_{\rm obs} > 10^5\ {\rm s}$, only an upper limit is
given, and the normal decay phase is not confirmed.  Thus, the
afterglow parameters are also hardly constrained from the observation.
We find that the theoretical LC (grey solid line) is basically
consistent with the observed upper limit for $\epsilon_e < 5 \times
10^{-4}$.  The right panel of Fig. \ref{fig:101225A} shows the
afterglow LCs in i (magenta) and r (green) bands.  Here we divide the
LCs into two phases by $t_{\rm obs} \sim 5 \ \rm day$.  In the earlier
phase, we suppose that the standard afterglow emissions (thin-dotted
lines) dominate, and find that the observed LCs are fitted with a
given set of the parameters, $p = 2.1$, $n = 0.1\ {\rm cm}^{-3}$,
$\epsilon_e = 5 \times 10^{-4}$, and $\epsilon_B = 0.05$.  In the
later phase, the HN-like bump dominates~\citep{2013arXiv1302.2352L},
and it can be fitted by the CFPEs (thin-dashed lines) with $A_{\rm
  V}^{\rm host} = 0.58$.

Although the uncertainties are relatively high, one can see that the
prompt and afterglow emissions of GRB 101225A are consistent
with the BSG model.

\subsection{GRB 121027A}

\begin{figure*}[!tb]
\begin{center}
\begin{tabular}{ll}
\resizebox{80mm}{!}{\includegraphics[]{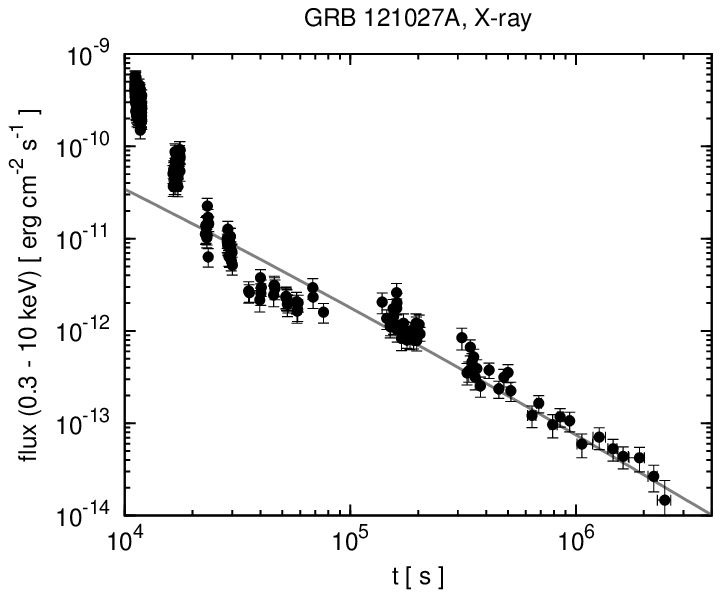}} &
\resizebox{80mm}{!}{\includegraphics[]{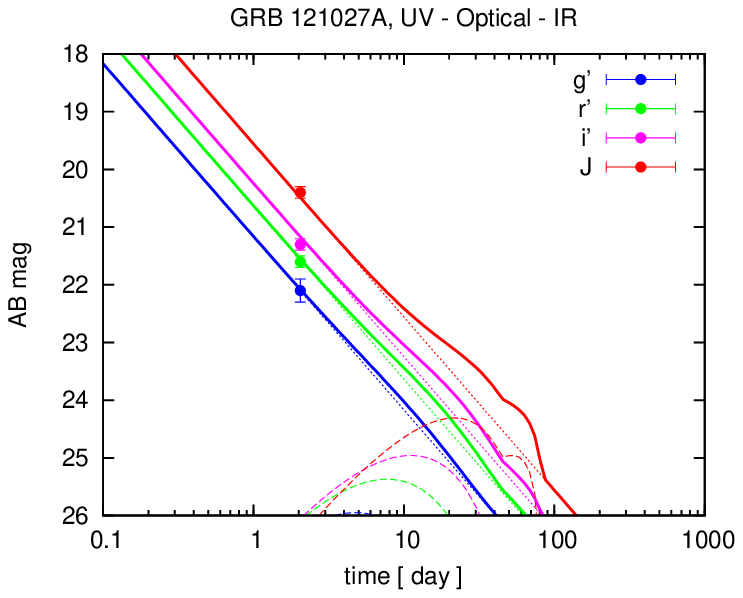}} \\
\end{tabular}
\caption{Same as Fig. \ref{fig:111209A}, but for GRB 121027A.  We find
  that the observed features are consistent with the BSG model, where
  we adopt the same progenitor model with Fig. \ref{fig:111209A}.  
  Since the observational data of the SN component are not available in this case, we need to fix  three parameters.
  If we give $\theta_{\rm j} = 12^{\circ}$, $n = 0.01\ {\rm cm}^{-3}$,
  and $A_{\rm V}^{\rm host} = 1.1$, for example, other parameters are determined
  from the observations as $\eta_{\rm j} = 1.24 \times 10^{-3}$,
  $\epsilon_{\gamma} = 0.1$, $p = 2.6$, $\epsilon_e = 0.05$, and $\epsilon_B = 8 \times 10^{-4}$
   (see Table \ref{tab:parameter}).  In
  the right panel, we suppose that the observed data at $\sim 2.05$
  day reflect the external shock emissions (thin-dotted lines).  The
  cocoon parameters are calculated as $E_{\rm c}(t_{\rm bo}) = 1.0
  \times 10^{53}\ {\rm erg}$ and $M_{\rm c}(t_{\rm bo}) =
  5.8\ M_{\odot}$.  We can predict that a SLSN-like bump derived from
  the CFPE (thin-dashed lines) might have been observed for $\sim
  20\mbox{-}80\ {\rm days}$ after the burst.}
\label{fig:121027A}
\end{center}
\end{figure*}

GRB 121027A consists of the prompt phase for $t_{\rm obs} < 200\ {\rm
  s}$ and the giant X-ray flare phase for $200\ {\rm s} < t_{\rm obs}
< 10^4\ {\rm s}$.  In this paper, we suppose that this giant X-ray
flare is also related to the central engine activity, and both
phases are due to the activity of the  jets.  We set the jet opening angle as
$\theta_{\rm j} = 12^{\circ}$, which is consistent with a constraint of $\theta_{\rm j} > 10^{\circ}$~\citep{2013arXiv1302.2352L}.
By adopting $\eta_{\rm j} = 1.24 \times 10^{-3}$ and
$\epsilon_{\gamma} = 0.1$, the observed prompt emission is roughly
reproduced as $E_{\rm iso} = 1.5 \times 10^{53}\ {\rm erg}$ and
$\delta t_\gamma \sim 15000\ \rm s$
\footnote{While we use the same values for $\theta_{\rm j}$ and $\eta_{\rm j}$ with those for GRB 111209A, the calculated duration $\delta t_\gamma \sim 15000\ \rm s$ is larger than that of GRB 111209A ($\delta t_\gamma \sim 9000\ \rm s$) calculated in \S \ref{subsec:111209A}. This is because GRB 121027A has higher redshift of $z = 1.773$ than that of GRB 111209A ($z = 0.677$).}.
The kinetic energy of the
relativistic ejecta and the cocoon parameters are given as $E_{\rm
  kin} = 1.4 \times 10^{54}\ {\rm erg}$, $E_{\rm c}(t_{\rm bo}) = 1.0
\times 10^{53}\ {\rm erg}$, and $M_{\rm c}(t_{\rm bo}) =
5.8\ M_{\odot}$, respectively (see Table \ref{tab:parameter}).

In Fig. \ref{fig:121027A}, we show the afterglow LCs in XRT (left
panel) and UV/optical/IR band (right panel).  In the right panel, each
color corresponds to g' (blue), r' (green), i' (magenta), and J (red)
band, respectively.  We find that the XRT LC for $t_{\rm obs} \gtrsim
2 \times 10^5 \ \rm s$ can be fitted well with $p = 2.6$ and
$\epsilon_e = 0.05$.  For UV/optical/IR band, the observed data have
been reported only at $t_{\rm obs} \sim 2.05\ {\rm day}$
\citep{Kruehler+12}.
Provided that the observed fluxes at $t_{\rm obs} \sim 2.05\ {\rm day}$ are attributed to the standard afterglow emissions, they are reproduced by setting $n = 0.01\ {\rm cm}^{-3}$, $\epsilon_B = 8 \times 10^{-4}$, and $A_{\rm V}^{\rm host} = 1.1$ (right panel).
The CFPEs are shown with thin-dashed lines, which slightly exceed the standard afterglow components in optical/IR bands at $t_{\rm obs} \sim 20\mbox{-}80\ \rm day$.
For this burst, the standard afterglow components are relatively bright due to the high acceleration efficiency of relativistic electrons, $\epsilon_e$ (see Table \ref{tab:parameter}).
So the CFPEs are almost buried compared to the standard afterglow components, although CFPEs are as bright as the observed SLSNe.
We predict that a SLSN-like bump might have appeared at $t_{\rm obs} \sim 20\mbox{-}80\ \rm day$ in the optical/IR afterglow light curve of GRB 121027A.

\section{Discussions}\label{sec:diss}

\subsection{Possible Subclass of SLSNe from Off-axis ULGRBs? }\label{subsec:slsn_cfpe}
In the previous section, we show that a ULGRB with a SLSN-like bump,
in particular GRB 111209A, can be interpreted by the collapsar-jet
scenario of BSG progenitors.  The prompt gamma-ray emissions
and the standard afterglow emissions are attributed to a relativistic
jet, while the SLSN-like bump comes from a non-relativistic cocoon fireball.  In this model, only the SLSN-like component may be seen, if
the observer locates along the off-axis direction from the ULGRB jet, due to
the relativistic beaming effect.  Then we can propose that a fraction
of SLSNe are originated from off-axis ULGRBs.  Indeed, the anticipated event
rate of such off-axis ULGRBs can be not much less than that of the
observed SLSNe \citep{Kashiyama_et_al_2013}.  This conjecture can be
tested, for example, by searching the orphan afterglow of a ULGRB associated with a SLSN.  After the relativistic jet is decelerated
enough, it begins to spread sideways to become a mildly-relativistic
spherical ejecta and afterglow emissions can be seen as an orphan
afterglow for off-axis observers afterward
it~\citep[e.g.,][]{1997ApJ...487L...1R, 1999ApJ...519L..17S}.
Thus, the
simultaneous and the follow-up observations in optical or radio bands
would play a key role for testing the proposed association.
Note that, by stacking the proposed SLSNe, we might also detect GeV-TeV
neutrinos, which are produced during the jet propagation inside BSG
progenitors~\citep{2013arXiv1306.2274M}.
 
\subsection{Parameter and Progenitor Dependence}\label{subsec:parameter_progenitor}
Here, we present the parameter dependence of our model.  First, let us
discuss the jet propagation inside a progenitor.  For a fixed jet
efficiency $\eta_{\rm j}$, a broader jet opening angle $\theta_{\rm
  j}$ gives a slower jet head.  If the jet head cannot reach the
surface of the progenitor before the engine stops, i.e., $\dot{M}(t)
\lesssim 10^{-3}\ M_{\odot}\ {\rm s}^{-1}$ (see Appendix
\ref{subsec:jet}), the jet fails to penetrate the progenitor.  In
particular, we find that for a fiducial jet efficiency $\eta_{\rm j} =
6.2 \times 10^{-4}$, jets fail to penetrate the progenitor BSG for
$\theta_{\rm j} \gtrsim 18^{\circ}$.  For twice energetic jets with
$\eta_{\rm j} = 1.24 \times 10^{-3}$, they fail to penetrate for $\theta_{\rm j}
\gtrsim 24^{\circ}$.  On the other hand, for a fixed $\theta_{\rm j}$,
there is a lower limit of $\eta_{\rm j}$ for the jet penetration.  For
the BSG progenitor with $\theta_{\rm j} = 12^{\circ}$, jets are
possible to penetrate the envelope for $\eta_{\rm j} \gtrsim 3.1 \times 10^{-4}$.

CFPEs are also affected by $\theta_{\rm j}$ and $\eta_{\rm j}$.  The
duration and the luminosity of CFPEs roughly scale with the internal
energy $E$, the baryon mass $M$, and the initial
cocoon-fireball radius $R_0$ as
\begin{equation}
\delta t_{\rm CFPE} \propto M^{3/4} E^{-1/4},
\label{eq:dura_cocoon}
\end{equation}
\begin{equation}
L_{\rm CFPE} \propto M^{-1} E R_{0}.
\label{eq:lumi_cocoon}
\end{equation}
Eqs. (3) and (4) can be easily derived if we remind the fact that
$ \delta t_{\rm CFPE}$ is determined by the equality between the 
photon diffusion time and the dynamical time $R(t)/v_{\rm sc}$ of the CFPE, while
the luminosity is derived from the photon diffusion  equation at  $\delta t_{\rm CFPE}$.
We find that for a fixed $\eta_{\rm j}$, $M$ increases with
$\theta_{\rm j}$, while $E$ hardly changes.  Therefore, one
can expect a longer and dimmer CFPE for a larger $\theta_{\rm j}$.  On
the other hand, for a fixed $\theta_{\rm j}$, $E$ increases
with $\eta_{\rm j}$, while $M$ hardly changes.  Therefore, one
can expect a shorter and brighter CFPE for a larger $\eta_{\rm j}$.

We also study the progenitor dependence of our model.
For that purpose, we fix the central engine parameters as $\theta_{\rm j} = 12^{\circ}$ and $\eta_{\rm j} = 1.24 \times 10^{-3}$, which are the same as those used in \S \ref{subsec:111209A} and are applied to the massive ($75\ M_{\odot}$) BSG progenitor.
We confirm that a lower mass BSG progenitor can reproduce the observed features of ULGRB prompt emissions.
For example, if we consider a BSG with pre-collapse mass $M_\ast = 40\ M_{\odot}$ and radius $R_\ast = 4.4 \times 10^{12}\ {\rm cm}$ \citep{2002RvMP...74.1015W}, the duration and the isotropic energy of GRB 111209A are roughly reproduced as $\delta t_\gamma \sim 6500 \ \rm s$ and $E_{\gamma, \rm iso} = 5.5 \times 10^{53} \ \rm erg$, respectively, with $\theta_{\rm j} = 12^{\circ}$, $\eta_{\rm j} = 1.24 \times 10^{-3}$, and $\epsilon_{\gamma} = 0.5$. 
Cocoon parameters are also calculated as $E_{\rm c}(t_{\rm bo}) = 5.6 \times 10^{52}\ {\rm erg}$ and $M_{\rm c}(t_{\rm bo}) = 7.0\ M_{\odot}$.
In this case, the CFPEs at $z=0.677$ have the peak magnitude of $\sim 23$ mag (with extinction of $A_{\rm V}^{\rm host} = 0.26$ mag) in J-band and the duration of $\sim 10\mbox{-}30$ day.
Note that for the massive BSG progenitor case, the peak magnitude of the CFPE is $\sim 22$ mag (with extinction of $A_{\rm V}^{\rm host} = 0.26$ mag) in J-band and the emission timescale is $\sim 20\mbox{-}40$ day (see Fig. \ref{fig:111209A}).
We find that while the duration of the CFPEs slightly differs, the luminosity of the CFPEs from the lower mass BSG  becomes several times lower than those from the more massive BSG progenitor.
Thus, we can infer the mass or radius of a BSG progenitor from the observation of CFPEs \citep[see also][]{Kashiyama_et_al_2013}.

We also find that WR progenitors cannot explain the observed SLSN-like bump in light of both the luminosity and the duration. 
A WR progenitor with ZAMS mass $40\ M_{\odot}$ and metallicity $Z_{\odot}$ has mass $M_\ast = 8.7\ M_{\odot}$ and radius $R_\ast = 2.3 \times 10^{10}\ {\rm cm}$ in the pre-collapse phase \citep{2002RvMP...74.1015W}.
We calculate the CFPEs from this WR progenitor by applying our model with $\theta_{\rm j} = 12^{\circ}$ and $\eta_{\rm j} = 1.24 \times 10^{-3}$, which give the cocoon parameters as $E_{\rm c}(t_{\rm bo}) = 7.3 \times 10^{51}\ {\rm erg}$ and $M_{\rm c}(t_{\rm bo}) = 0.33\ M_{\odot}$, respectively.
We find that the peak magnitude of the CFPEs at $z=0.677$ becomes $\sim 27$ mag (with extinction of $A_{\rm V}^{\rm host} = 0.26$ mag) in J-band, which is almost 100 times dimmer than those for the massive ($75\ M_{\odot}$) BSG progenitor case.
The light curve peak is reached on $\sim 3\ {\rm day}$ in J-band, which is much earlier than that of the BSG progenitor case.
From these results, we conclude that bright CFPEs may be an unique characteristic of GRBs from BSG progenitors \citep[see also][]{Kashiyama_et_al_2013}.

\subsection{CFPEs from Population III GRBs}\label{subsec:popIII_GRB}
An interesting application of the BSG model is to Population III (Pop
III) stars, metal-free stars formed dominantly at high redshifts.  Pop
III stars are considered to become massive BSGs in their pre-collapse
phase because of their metal-free envelopes, although the typical mass
has a large theoretical uncertainty, i.e., $\sim 10-10^3 \ M_{\rm
  \odot}$ \citep[e.g.,][]{2002Sci...295...93A, 2002ApJ...564...23B,
  2004ApJ...603..383T, 2008ApJ...681..771M, 2011Sci...334.1250H}.
This implies that Pop III GRBs can be seen as ULGRBs, and they potentially dominate at higher redshifts if
the BSG model holds.  For broad mass range of Pop III BSG progenitors,
many authors have argued that both prompt and standard afterglow signals of Pop III
GRBs would be detectable using future facilities \citep{2010ApJ...715..967M, 2010MNRAS.402L..25K,
  2011ApJ...726..107S, 2012ApJ...754...85N, 2012ApJ...759..128N,
  Kashiyama_et_al_2013}.

As we have discussed so far, bright CFPEs may be common
characteristics of BSG GRBs. In addition to the prompt and standard
afterglow emissions, CFPEs from Pop III GRBs will be detectable even
at $z \gtrsim 10$ with future facilities like {\it JWST} NIRCam
\citep{Kashiyama_et_al_2013}, which will provide us valuable
information about the star formation history in the high-$z$ universe.

\subsection{Other Scenarios of ULGRBs}\label{subsec:other_scenario}
Finally, we here discuss other possible scenarios proposed to explain ULGRBs. 

For GRB 101225A, two different scenarios were proposed just after the
discovery.  One is the tidal disruption of a minor body by a galactic
neutron star \citep{2011Natur.480...69C}, and the other is related to
the stellar merger of a neutron star with a helium star at $z =
0.33$~\citep{2011Natur.480...72T}.  Now we know that the observed
ULGRBs are extragalactic origin, the former galactic scenario is
unlikely.  On the other hand, the latter scenario predicts that
afterglow emissions are suppressed as in GRB 101225A, and thus cannot
be directly applied to GRB 111209A and 121027A.

In collapsar-jet scenario, a longer activity of a central engine can
be realized by implementing a fall-back
accretion~\citep{2012MNRAS.419L...1Q}.  In fact,
\cite{2013ApJ...767L..36W} divide GRB 121027A into a prompt phase with
a duration of $\sim 200\ {\rm s}$ and an X-ray-flare phase with $\sim
10^4\ {\rm s}$, and showed that the X-ray flare with a sharp rising
can be explained by a fall-back accretion of falling back mass $M_{\rm fb} \sim
0.9-2.6\ M_{\odot}$ from the radius $r_{\rm fb} \sim 3.5 \times 10^{10}\ {\rm cm}$.
Although our model does not take into account the effect of fall-back
accretion, the BSG model itself is compatible with this
interpretation, since the fall-back accretion can naturally occur in
massive BSGs.

Shock-breakout scenario can also reproduce the observed duration of
ULGRBs by implementing a large breakout radius, $c\ \delta t_\gamma
\sim 3 \times 10^{14}\ (\delta t_\gamma / 10^{4}\ {\rm s})\ {\rm cm}$
\citep{1974ApJ...187..333C, 1978ApJ...223L.109K, 1992ApJ...393..742E,
  1999ApJ...510..379M}.  However, shock-breakout emissions would show
relatively smooth light curves and quasi-thermal
spectra~\citep[e.g.,][]{2007ApJ...667..351W, 2012ApJ...747...88N},
which is not the case for the observed ULGRBs.

Tidal disruptions of stars by supermassive black holes can also give
long lasting emissions~\citep{2011Natur.476..421B,
  2011Sci...333..203B}.  However, the observed tidal disruption events
have isotropic luminosities $\lesssim 10^{48}\ {\rm erg}\ {\rm
  s}^{-1}$, which are orders-of-magnitude dimmer than ULGRBs ($\gtrsim
10^{49}\ {\rm erg}\ {\rm s}^{-1}$).  Also, this scenario predicts a
prompt light curve with temporal dependence $\propto
t^{-5/3}$~\citep{1988Natur.333..523R, 1989IAUS..136..543P}, which is
inconsistent with the observed ULGRBs.

\section{Summary}\label{sec:sum}
ULGRBs have durations of $\sim 10^4\ {\rm s}$, which is much longer
than that of LGRBs ($\sim 30\ {\rm s}$).  The observed afterglow light
curves show SLSN-like bumps, which are $\lesssim 10$ times brighter than
a GRB-associated hypernova, SN 1998bw.  Such long durations are naturally
explained, if we consider BSG collapsars rather than WR collapsars,
since the accretion of a massive hydrogen envelope activates the
central engine for a long duration.  In the BSG model, ULGRBs necessarily
accompany SLSN-like transients in afterglow phase, as shown in
\cite{Kashiyama_et_al_2013}.  They considered photospheric emissions
from the expanding cocoon fireball, which is formed during the jet
propagation in the progenitor and breaks out the progenitor along with
the jet.  In this paper, we refine our previous model for
cocoon-fireball photospheric emissions (CFPEs), and interpret the
observed SLSN-like bump as the CFPE.  For GRB 111209A and GRB 101225A,
we find that if we have enough observations, i.e., the duration,
isotropic energy, X-ray and UV/optical/IR afterglow light curves, and
SLSN-like bump, the observed features of a ULGRB and SLSN-like bump
are reproduced quite well, for a given set of the progenitor model;
the jet opening angle, and the ambient gas density.  For GRB 121027A,
UV/optical/IR afterglow data after $t_{\rm obs} = 2.05\ {\rm day}$ are
not published yet, and we predict that SLSN-like bump might have been
observed for $\sim 20\mbox{-}80$ days after the burst.

\section*{Acknowledgements}
We thank K. Ioka, T. Enoto, and T. Sakamoto for fruitful discussions
and comments.
We also thank P. M{\'e}sz{\'a}ros, B.B. Zhang, P. Veres, and K. Murase for valuable discussions.
This work is supported in part by the Grant-in-Aid from
the Ministry of Education, Culture, Sports, Science and Technology
(MEXT) of Japan, Nos. 23540305 (TN), 24103006 (TN), 23840023
(YS), 25103511 (YS), JSPS Postdoctoral Fellowship for Research
Abroad (KK), and NASA NNX13AH50G (KK).

\appendix
\section{Calculation Methods}\label{app}
In this appendix, we show the calculation method we use for
evaluating electromagnetic emissions associated with BSG GRBs.  Our
method is based on the collapsar-jet scenario.  Once a progenitor
model and a set of phenomenological parameters are fixed, we can
calculate physical quantities of the prompt emission, the afterglow
emission, and the cocoon-fireball photospheric emission (CFPE) in a self-consistent manner.

\subsection{Jet-cocoon Formation inside Progenitors}\label{subsec:jet}
First, we model jet-cocoon formation within progenitors following our
previous studies
\citep{2011ApJ...726..107S,2012ApJ...759..128N,Kashiyama_et_al_2013}.
In this paper, we consider massive progenitors with $M \gtrsim
40\ M_{\odot}$, which will collapse directly into a black hole (BH) without
significant mass ejections \citep{2003ApJ...591..288H}.  We assume
that a bi-polar relativistic jet is launched when the mass of the
central BH becomes $3\ M_{\odot}$.  The jet luminosity is proportional
to the mass accretion rate onto the central BH,
\begin{equation}
L_{\rm j}(t) = \eta_{\rm j} \dot{M}(t) c^2. 
\label{eq:jet}
\end{equation}
This can be justified for jets driven by magneto-hydrodynamic
mechanisms \citep[e.g.,][]{2010MNRAS.402L..25K}.  Here, $t$ is the
time since the central engine becomes active in the GRB rest frame.
The mass accretion rate $\dot{M}(t)$ can be estimated as
\begin{equation}
\dot{M} = \alpha \frac{d M_r}{d t_{\rm ff}(r)},
\label{eq:mdot}
\end{equation} 
where $M_r$ is the mass coordinate, and $\alpha$ represents the effect
of disk accretion, which is set as $\alpha = 1$ throughout this paper
\citep{2008MNRAS.388.1729K}.

The velocity of the jet head can be obtained from the pressure balance
at the interface of the jet and the stellar
envelope~\citep{2003MNRAS.345..575M},
\begin{gather}
\beta_{\rm h}(t) =  \frac{\beta_{\rm j}}{1+ \Tilde{L}(t)^{-1/2}}, \\
\Tilde{L}(t) = \frac{L_{\rm j}(t-r_{\rm h}/(\beta_{\rm j} c))}{\Sigma_{\rm j}(t) \rho_{\ast}(r_{\rm h}) c^3}, \notag
\label{eq:tildeL}
\end{gather}
where $\beta_{\rm j} \approx 1$ is the velocity of the jet and 
$\rho_{\ast}(r)$ is the density of the stellar envelope.  The radius of the jet head
is obtained from $r_{\rm h}(t) = \int^t \beta_{\rm h}(t') c\ dt'$ and
$\Sigma_{\rm j}(t) = \pi r_{\rm h}^2(t) \theta_{\rm j}^2$ represents
the cross section of the jet head where $\theta_{\rm j}$ is the jet opening angle.  
We regard that the jet breaks out
the envelope when $r_{\rm h}(t_{\rm bo}) = R_{\ast}$ where $t_{\rm bo}$ is the time of the jet break out.  A successful
GRB is expected if the central engine is still active after the
breakout time, which corresponds to $\dot{M}(t) \gtrsim 10^{-3}
M_{\odot} \ \rm s^{-1}$~\citep[e.g.,][]{Chen_Beloborodov_2007} at $t >
t_{\rm bo}$ since the neutrino cooling is not effective for $\dot{M}(t) <10^{-3}
M_{\odot} \ \rm s^{-1}$.  For BSG progenitors, we showed that jets can penetrate
the progenitor envelopes irrespective of their masses
\citep{2011ApJ...726..107S, 2012ApJ...759..128N,Kashiyama_et_al_2013}.

As far as the jet head is non-relativistic, i.e., $\beta_{\rm h} \ll
1$, shocked matter at the jet head will spread out sideways, and form
a cocoon.  The cocoon expands in the stellar envelope with a
transverse velocity of
\begin{equation}
\beta_{\rm c}(t) \sim \sqrt{\frac{E_{\rm c}(t)}{3 \rho_{\ast}(r_{\rm h}) c^2 V_{\rm c}(t)}}, 
\label{eq:betac}
\end{equation}
which is obtained from the pressure balance at the interface of the
cocoon and the stellar envelope~\citep{2003MNRAS.345..575M}\footnote{\cite{2003MNRAS.345..575M} represents the cocoon pressure as $P_{\rm c} \sim E_{\rm c} / 3 \rho_\ast V_{\rm c}$, which is a typo. Correctly, it should be $P_{\rm c} \sim E_{\rm c}/3V_{\rm c}$.}. 
We assume that the cocoon has a conical shape with a height of $r_{\rm
  h}(t)$ and a circular radius of $r_{\rm c}(t) = \int^t \beta_{\rm
  c}(t')c\ dt'$ at the bottom.  Then, the volume of the cocoon $V_{\rm
  c}$ can be estimated as $V_{\rm c}(t) = \pi r_{\rm c}(t)^2 r_{\rm
  h}(t) / 3$.  As it expands, cocoon also loads the stellar material
along the direction of motion and the mass loaded in the cocoon can be evaluated
from
\begin{equation}
M_{\rm c}(t) = \frac{r_{\rm c}(t)^2}{4 r_{\rm h}(t)^2} \int^{r_{\rm h}(t)} 4 \pi r^2 \rho_{\ast}(r) dr.
\label{eq:mass_c}
\end{equation}
Before the jet breakout, most of the jet energy is stored in the
cocoon and the cocoon energy can be described as $E_{\rm c}(t) =
\int^t L_{\rm j}(t'-r_{\rm h}/c) dt'$.

In summary, we can calculate the jet breakout time $t_{\rm bo}$,
the cocoon energy $E_{\rm c}(t_{\rm bo})$ and  the mass of the cocoon $M_{\rm
  c}(t_{\rm bo})$ by fixing the progenitor model and the central
engine parameters $(\theta_{\rm j}, \eta_{\rm j})$.  In this paper, we
assume that both $\eta_{\rm j}$ and $\theta_{\rm j}$ are constants
for simplicity~\citep[but
  see e.g.,][]{Kawanaka_et_al_2013,Mizuta_Ioka_2013}.

\subsection{Prompt Emission}\label{subsec:prompt}
After the jet breakout, a fraction of the jet energy will be
dissipated and radiated as prompt gamma-rays.  Since the mechanism is
still highly uncertain, we here do not discuss the energy spectrum
\citep[see][for discussion about the spectrum of prompt emissions]{2012ApJ...759..128N}.  Instead, we estimate the
isotropic energy $E_{\gamma, \rm iso}$ and the prompt duration $\delta t_\gamma$, and
compare them with the observed ones.

In general, the dissipation radius is larger than the progenitor
radius, and the prompt emission starts at $t \sim t_{\rm bo}$.
Following the results of e.g., \cite{Chen_Beloborodov_2007}, we
suppose that the prompt emission ends when the mass accretion rate
becomes smaller than the critical value $\dot{M}(t_{\rm fin}) \sim
10^{-3}\ M_{\odot} \ \rm s^{-1}$.  Hence, one can evaluate the
duration of the prompt emission as
\begin{equation}
\delta t_\gamma = (t_{\rm fin} - t_{\rm bo}) (1+z), 
\end{equation}
and the isotropic energy as 
\begin{equation}
E_{\gamma, \rm iso} = \int_{t_{\rm bo}}^{t_{\rm fin}} L_{\rm iso}(t') dt',
\label{eq:e_giso}
\end{equation}
where $L_{\rm iso}(t) = \epsilon_{\gamma} (4/\theta_{\rm j}^2) L_{\rm
  j}(t)$ is the isotropic luminosity of the prompt emission, and
$\epsilon_\gamma$ is the radiation efficiency.

In summary, we fix ($\theta_{\rm j}$, $\eta_{\rm j}$, $\epsilon_\gamma$) to reproduce $E_{\gamma, \rm iso}^{\rm obs}$ and $\delta t_\gamma^{\rm obs}$ of ULGRBs.
These parameters have been inferred for LGRBs both observationally and theoretically.
The typical value of $\theta_{\rm j}$ is estimated as $\theta_{\rm j} \sim 5^{\circ}$ for bursts with jet-break signature in the afterglow light curve~\citep{2004ApJ...616..331G, 2006ApJ...650..261S}.
From the prompt and afterglow observations, $\epsilon_\gamma$ is estimated as $\epsilon_\gamma \sim 0.01\mbox{-}1$~\citep{2007ApJ...655..989Z}.  The observed LGRBs typically have
$E_{\gamma, \rm iso}^{\rm obs} \sim 10^{52}\mbox{-}10^{54}\ \rm erg$
and $\delta t^{\rm obs} \sim 10\mbox{-}100\ \rm s$.
\cite{2011ApJ...726..107S} argued that these typical features of LGRBs can be reproduced for a WR progenitor, if we apply our prescription, setting $\theta_{\rm j} \sim 5^{\circ}$, $\epsilon_\gamma \sim 0.01\mbox{-}1$, and $\eta_{\rm j} = 6.2 \times 10^{-4}$.
\cite{Kawanaka_et_al_2013} theoretically estimated the jet efficiency as $\eta_{\rm j} \sim 10^{-4}\mbox{-}10^{-3}$.
In this paper, we take $\eta_{\rm j} = 6.2 \times 10^{-4}$ as a fiducial value, and change its value as $\eta_{\rm j} \sim 10^{-4}\mbox{-}10^{-3}$ for fitting.

\subsection{Afterglow Emission}\label{subsec:afg}
We calculate the afterglow emissions based on the standard external
shock model \citep{1998ApJ...497L..17S}.  The relativistic jet finally
decelerates in the interstellar medium, where a fraction of electrons
is accelerated to relativistic energies at the forward shock, and
emits synchrotron radiation in magnetic fields amplified by the
shock.  For the observed ULGRBs, the normal decay phase of the
afterglow starts at $t_{\rm obs} \sim 10^5\ {\rm s}$, which
corresponds to the slow cooling phase, and the jet break is not
confirmed until $\sim 2 \times 10^6$\ s \citep{2013arXiv1302.2352L}.
If we follow Eqs. (8) and (11) in \cite{1998ApJ...497L..17S}, the light curves can be modeled as
\begin{equation}
F_{\nu} \sim \begin{cases}
0.061 \ f(p)^{p-1} \epsilon_{e, -1}^{p-1} \epsilon_{B, -2}^{\frac{p+1}{4}} E_{{\rm kin}, 53}^{\frac{p+3}{4}} n^{1/2} D_{28}^{-2} (1+z)^{\frac{3-p}{2}} t_4^{-\frac{3p-3}{4}} \left(\frac{\nu}{10^{14}\ {\rm Hz}}\right)^{\frac{1-p}{2}}\ {\rm Jy} & (\nu_{\rm m} < \nu < \nu_{\rm c}), \\
\\
9.8 \times 10^{-4}\  f(p)^{p-1} \epsilon_{e, -1}^{p-1} \epsilon_{B, -2}^{\frac{p-2}{4}} E_{{\rm kin}, 53}^{\frac{p+2}{4}} n^{0} D_{28}^{-2} (1+z)^{\frac{2-p}{2}} t_4^{-\frac{3p-2}{4}} \left(\frac{\nu}{10^{16}\ {\rm Hz}}\right)^{-\frac{p}{2}}\ {\rm Jy} & (\nu_{\rm c} < \nu),
\label{eq:sp_light_curve}
\end{cases}
\end{equation}
with the characteristic frequency and the cooling frequency of
synchrotron radiation
\begin{equation}
\nu_{\rm m} \sim  4.6 \times 10^{13}\ f(p)^2 \epsilon_{e, -1}^{2} \epsilon_{B, -2}^{1/2} E_{{\rm kin}, 53}^{1/2} (1+z)^{-1} t_4^{-3/2} {\rm Hz},
\label{eq:nu_m}
\end{equation}
and
\begin{equation}
\nu_{\rm c} \sim  2.6 \times 10^{15}\ \epsilon_{B, -2}^{-3/2} E_{{\rm kin}, 53}^{-1/2}  n^{-1} (1+z)^{-1} t_4^{-1/2} {\rm Hz}, 
\label{eq:nu_c}
\end{equation}
where we use $Q_x = Q / 10^{x}$ in CGS units, and $\nu$ is the
observed photon frequency. Here, $p$ is the power law index of the
non-thermal electrons, $f(p)=3(p-2)/(p-1)$, $\epsilon_{e}$ is the
electron-acceleration efficiency, $\epsilon_{B}$ is the amplification
efficiency of magnetic field, and $n$ is the ambient matter density.
$D$ and $z$ are the luminosity distance and the red shift of the
source, respectively.  Using $E_{\gamma, \rm iso}$ and
$\epsilon_{\gamma}$ introduced in \S \ref{subsec:prompt}, the kinetic
energy of the relativistic ejecta is calculated as $E_{\rm kin} =
E_{\gamma, \rm iso} (1 - \epsilon_{\gamma})/\epsilon_{\gamma}$.

In X-ray and UV/optical/IR bands, the extinction both in the Milky Way
(MW) and the host galaxy is important.  For X-ray absorption, we use
the cross section of the photoelectric absorption per HI atom shown in
\cite{1983ApJ...270..119M}, assuming solar abundance.  The measured HI
column density of the MW is $N_{\rm H}^{\rm MW} = 1.5 \times
10^{20}\ {\rm cm}^{-2}$ \citep{2005A&A...440..775K}.  The HI column
densities of the host galaxies are measured as $N_{\rm H}^{\rm host} =
1.2 \times 10^{21}\ {\rm cm}^{-2}$ for GRB 101225
\citep{2013arXiv1302.2352L}, $N_{\rm H}^{\rm host} = 1.9 \times
10^{21}\ {\rm cm}^{-2}$ for GRB 111209 \citep{2013arXiv1302.2352L},
and $N_{\rm H}^{\rm host} = 1.3 \times 10^{22}\ {\rm cm}^{-2}$ for GRB
121027 \citep{2013arXiv1302.4876P}, respectively.
In UV/optical/IR bands, we refer the MW extinction for each burst to \cite{2013arXiv1302.2352L}.
The host extinction in each band can be calculated from the formula
\begin{equation}
A_{\lambda}^{\rm host} = \xi (\lambda) \left(1+\frac{1}{R_V} \right)A_V^{\rm host},
\label{eq:A_lambda}
\end{equation}
where $\xi (\lambda) \equiv A_{\lambda}^{\rm host}/ A_{B}^{\rm host}$, and $R_V \equiv A_V^{\rm host} / E_{B\mbox{-}V}$ is the ratio of total-to-selective extinction.
Three types of extinction curves are studied in \cite{1992ApJ...395..130P} for the MW, Large Magellanic Cloud, and Small Magellanic Cloud (SMC), and they are used in evaluating the extinction of other galaxies with the parameter, $A_V^{\rm host}$.
We apply the SMC type extinction law for ULGRB host, since they are blue compact galaxies.
We refer the fitting formula of $\xi (\lambda)$ to Eq. (20) and Table 4 in \cite{1992ApJ...395..130P}, and the value of $R_V=2.93$ to Table 2 in \cite{1992ApJ...395..130P}. Thus, extinctions are parameterized only by $A_{\rm V}^{\rm host}$.

In summary, we additionally introduce 5 parameters ($p$,
$\epsilon_{e}$, $\epsilon_{B}$, $n$, $A_{\rm V}^{\rm host}$) to fit
the standard afterglow component in XRT ($0.3\mbox{-}10 \ \rm keV$)
and UV/optical/IR bands.  Practically, there are $4$ constraints on
these parameters from the observations.  For some GRBs, the range of 4
parameter ($p$, $\epsilon_{e}$, $\epsilon_{B}$, $n$) values are
obtained from the afterglow observations as $p \sim 1.4\mbox{-}2.8$,
$\epsilon_{e} \sim 4 \times 10^{-3}\mbox{-}0.1$, $\epsilon_{B} \sim 4
\times 10^{-5}\mbox{-}0.07$, and $n \sim 0.01\mbox{-}100\ {\rm
  cm}^{-3}$, respectively \citep[e.g.,][]{2002ApJ...571..779P}.  From
observation, the V-band extinction takes values of  $A_{\rm V}^{\rm host} \sim 0.1\mbox{-}
3.0$~\citep{2013arXiv1303.4743C,2013arXiv1303.6924Z}.

\subsection{Cocoon-Fireball Photospheric Emission}\label{subsec:cocoon}
When jets penetrate the progenitors, hot-plasma cocoons also come out.
Such cocoons are radiation-dominated and non-relativistic, i.e.,
$E_{\rm c}(t_{\rm bo})/M_{\rm c}(t_{\rm bo}) c^2 <
1$~\citep{Kashiyama_et_al_2013}.  They first expand almost
adiabatically, and then emit quasi-thermal photons from the
photosphere, where the Thomson optical depth $\tau_{\rm T}$ becomes
unity.  Since cocoon fireballs would load hydrogens from BSG
envelopes, these processes are quite similar to those of shock-heated
ejecta responsible for Type IIP SNe, which have been well modeled
\citep{1980ApJ...237..541A, 1993ApJ...414..712P, 2009ApJ...703.2205K}.
Here, we calculate CFPEs following \cite{1980ApJ...237..541A} and
\cite{1993ApJ...414..712P}.

\cite{1980ApJ...237..541A} and \cite{1993ApJ...414..712P} analytically
formulated photospheric emissions from spherically expanding
shock-heated ejecta.  There are $4$ input parameters; the initial
ejecta radius $R_0$, the internal energy $E \equiv E_{\rm c}(t_{\rm bo})$, the baryon mass $M \equiv M_{\rm c}(t_{\rm bo})$,
and the expansion velocity scale $v_{\rm sc}$.  Here, we set the initial conditions
for cocoon fireballs at $R_0 \sim 2 R_\ast$, where roughly a half the
internal energy is transferred to the kinetic energy and cocoon
fireballs become almost spherical;
\begin{equation}
E_{\rm th, 0} = E/2,
\end{equation}
\begin{equation}
v_{\rm sc} = \sqrt{10E/3M}. 
\end{equation}

Beyond the saturation radius, $R(t) \sim v_{\rm sc} \times t > R_0$, we consider a
homologous expansion of cocoon fireballs \citep{1980ApJ...237..541A,
  1993ApJ...414..712P};
\begin{equation}
r = x \times R(t),
\label{eq:EOM}
\end{equation} 
\begin{equation}
v(r, t) = x \times v_{\rm sc}. 
\end{equation} 
We assume uniform densities of $\rho(t) = \rho_0 (R(t)/R_0)^{-3}$ with
$\rho_0 = 3M/4\pi R_0{}^3$.  The temperature structure inside the
fireball, $T(r,t)$, can be obtained by solving the equation below;
\begin{equation}
\frac{\partial e}{\partial t} + P \frac{\partial}{\partial t} \left(\frac{1}{\rho} \right) = - \frac{\partial L}{\partial m_r},
\label{eq:energy_eq}
\end{equation} 
where $e(t) = a T^4(t) / \rho(t)$ is the specific internal energy,
$P(t) = a T^4(t) / 3$ is the radiation pressure, $L = - (4 \pi r^2
ac/3 \kappa_{\rm T} \rho) \times \partial T^4/\partial r$ represents
the radiative energy loss, $m_r$ is the mass coordinate, and $a = 7.56
\times10^{-15}\ {\rm erg}\ {\rm cm}^{-3}\ {\rm K}^{-4}$ is the
radiation constant.  For the opacity of cocoon fireballs, $\kappa$, we
simply assume a step function;
\begin{equation}
\kappa = \left\{
\begin{array}{ll}
\kappa_{\rm T} = 0.34\ {\rm cm}^2\ {\rm g}^{-1} & T > T_{\rm ion}, \\
0 & T < T_{\rm ion}.
\end{array}
\right.
\label{eq:opacity}
\end{equation}
Here, $\kappa_{\rm T}$ is the Thomson opacity, and $T_{\rm ion}
\approx 6000 \ \rm K$ is the hydrogen recombination temperature in the cocoon
fireball.  For $T < T_{\rm ion}$, the recombination of hydrogen atoms
occurs and the opacity becomes effectively zero.

\cite{1980ApJ...237..541A} and \cite{1993ApJ...414..712P} solved Eq. \eqref{eq:energy_eq} with Eq. \eqref{eq:opacity} where the photosphere radius evolves  as 
\begin{equation}
R_{\rm ph}(t) \sim \left\{
\begin{array}{ll}
R(t), & (t_{\rm e} \ll t < t_{\rm i}), \\
R_{\rm ion}(t), & (t_{\rm i} < t). 
\end{array}
\right.
\label{eq:R_ph}
\end{equation}
Here, $t_{\rm e} = R_0 / v_{\rm sc}$, and $t_{\rm i}$ corresponds to the time
at which the effective temperature of the photosphere becomes equal to the hydrogen recombination temperature $T_{\rm ion} \approx 6000\ \rm K$ as shown below.
Afterward it, the recombination front appears in the cocoon fireball.
The recombination front, $R_{\rm ion}(t)$, evolves as
\begin{equation}
R_{\rm ion}(t)^2 = v_{\rm sc}^2 \left[ t_{\rm i} t \left(1+\frac{t_{\rm i}^2}{3 t_{\rm a}^2} \right) - \frac{t^4}{3 t_{\rm a}^2}\right].
\label{eq:R_ion}
\end{equation}
with $t_{\rm a} = \sqrt{2 t_{\rm d} t_{\rm e}}$ and the photon diffusion  time $t_{\rm d} \equiv 9 \kappa_{\rm T}
M/(4 \pi^3 c R_0)$.
The bolometric luminosity of the photospheric emission is described as
\begin{equation}
L(t) = \left\{
\begin{array}{ll}
(E_{\rm th, 0}/t_{\rm d}) \exp\left(- t^2 / t_{\rm a}^2 \right), & (t_{\rm e} \ll t < t_{\rm i}), \\
4 \pi R_{\rm ion}(t)^2 \sigma_{\rm SB} T_{\rm ion}^4 & (t_{\rm i} < t),
\end{array}
\right.
\label{eq:luminosity}
\end{equation}
which gives the effective temperature of the photosphere as  
\begin{equation}
T_{\rm eff}(t) = \left\{
\begin{array}{ll}
T_{\rm eff}(0) \exp\left(-t^2/4 t_{\rm a}^2\right) \left(t_{\rm e}/ t\right)^{1/2}, & (t_{\rm e} \ll t < t_{\rm i}), \\
T_{\rm ion} & (t_{\rm i} < t).
\end{array}
\right.
\label{eq:T_eff}
\end{equation}
Here, $T_{\rm eff}^4(0) \equiv E_{\rm th, 0}/(4 \pi R_0^2 \sigma_{\rm SB} t_{\rm d})$
and $\sigma_{\rm SB}$ is the Stefan-Boltzmann constant.  The
transition time $t_{\rm i}$ can be iteratively determined from the
condition $T_{\rm eff}(t_{\rm i}) = T_{\rm ion}$.

Now that the evolution of the photospheric radius and its temperature
is determined, one can calculate the CFPE as
\begin{equation}
F_{\lambda_{\rm obs}}(t)\ d \lambda_{\rm obs} = \pi B_{\lambda}(T_{\rm eff}(t))\ d \lambda \ \frac{R_{\rm ph}(t)^2}{D^2},
\label{eq:lambda}
\end{equation}
where $B_{\lambda}(T) = 2 h c^2 / \lambda^5 (\exp(hc/\lambda k_{\rm B}
T) - 1)^{-1}$, $h$ is the Planck constant and $k_{\rm B}$ is the
Boltzmann constant.

\end{document}